\documentclass[a4paper]{article}
\usepackage[english]{babel}
\usepackage{amsmath,amsfonts,amssymb}
\usepackage{graphicx}

\begin{document}

\relax
\renewcommand{\theequation}{\arabic{section}.\arabic{equation}}

\def\be{\begin{equation}}
\def\ee{\end{equation}}
\def\bs{\begin{subequations}}
\def\es{\end{subequations}}
\def\calm{{\cal M}}
\def\calk{{\cal K}}
\def\Hc{{\cal H}}
\def\lx{\lambda}
\def\sx{\sigma}
\def\ex{\epsilon}
\def\Lx{\Lambda}

\newcommand{\bett}{\tilde{\beta}}
\newcommand{\alpht}{\tilde{\alpha}}
\newcommand{\dv}{\delta{v}}
\newcommand{\dvvec}{\delta\vec{v}}
\newcommand{\phit}{\tilde{\phi}}
\newcommand{\Phit}{\tilde{\Phi}}
\newcommand{\Gt}{\tilde{G}}
\newcommand{\mt}{\tilde{m}}
\newcommand{\Mt}{\tilde{M}}
\newcommand{\nt}{\tilde{n}}
\newcommand{\Nt}{\tilde{N}}
\newcommand{\Bt}{\tilde{B}}
\newcommand{\Rt}{\tilde{R}}
\newcommand{\rt}{\tilde{r}}
\newcommand{\mut}{\tilde{\mu}}
\newcommand{\mub}{\bar{\mu}}
\newcommand{\vrm}{{\rm v}}
\newcommand{\tl}{\tilde t}
\newcommand{\ttt}{\tilde T}
\newcommand{\rhot}{\tilde \rho}
\newcommand{\ptt}{\tilde p}
\newcommand{\drho}{\delta \rho}
\newcommand{\dpp}{\delta p}
\newcommand{\dphi}{\delta \phi}
\newcommand{\drhot}{\delta {\tilde \rho}}
\newcommand{\dchi}{\delta \chi}
\newcommand{\A}{A}
\newcommand{\B}{B}
\newcommand{\mmu}{\mu}
\newcommand{\mnu}{\nu}
\newcommand{\ii}{i}
\newcommand{\jj}{j}
\newcommand{\jl}{[}
\newcommand{\jr}{]}
\newcommand{\ml}{\sharp}
\newcommand{\mr}{\sharp}

\newcommand{\da}{\dot{a}}
\newcommand{\db}{\dot{b}}
\newcommand{\dn}{\dot{n}}
\newcommand{\dda}{\ddot{a}}
\newcommand{\ddb}{\ddot{b}}
\newcommand{\ddn}{\ddot{n}}
\newcommand{\pa}{a^{\prime}}
\newcommand{\pn}{n^{\prime}}
\newcommand{\ppa}{a^{\prime \prime}}
\newcommand{\ppb}{b^{\prime \prime}}
\newcommand{\ppn}{n^{\prime \prime}}
\newcommand{\fda}{\frac{\da}{a}}
\newcommand{\fdb}{\frac{\db}{b}}
\newcommand{\fdn}{\frac{\dn}{n}}
\newcommand{\fdda}{\frac{\dda}{a}}
\newcommand{\fddb}{\frac{\ddb}{b}}
\newcommand{\fddn}{\frac{\ddn}{n}}
\newcommand{\fpa}{\frac{\pa}{a}}
\newcommand{\fpb}{\frac{\pb}{b}}
\newcommand{\fpn}{\frac{\pn}{n}}
\newcommand{\fppa}{\frac{\ppa}{a}}
\newcommand{\fppb}{\frac{\ppb}{b}}
\newcommand{\fppn}{\frac{\ppn}{n}}
\newcommand{\at}{\tilde{\alpha}}
\newcommand{\pt}{\tilde{p}}
\newcommand{\Ut}{\tilde{U}}
\newcommand{\phidot}{\dot{\phi}}
\newcommand{\rhb}{\bar{\rho}}
\newcommand{\pb}{\bar{p}}
\newcommand{\pbb}{\bar{\rm p}}
\newcommand{\kt}{\tilde{k}}
\newcommand{\kb}{\bar{k}}
\newcommand{\wt}{\tilde{w}}

\newcommand{\dA}{\dot{A_0}}
\newcommand{\dB}{\dot{B_0}}
\newcommand{\fdA}{\frac{\dA}{A_0}}
\newcommand{\fdB}{\frac{\dB}{B_0}}

\def\be{\begin{equation}}
\def\ee{\end{equation}}
\def\bs{\begin{subequations}}
\def\es{\end{subequations}}
\newcommand{\een}{\end{subequations}}
\newcommand{\ben}{\begin{subequations}}
\newcommand{\beq}{\begin{eqalignno}}
\newcommand{\eeq}{\end{eqalignno}}

\def \lta {\mathrel{\vcenter
     {\hbox{$<$}\nointerlineskip\hbox{$\sim$}}}}
\def \gta {\mathrel{\vcenter
     {\hbox{$>$}\nointerlineskip\hbox{$\sim$}}}}

\def\g{\gamma}
\def\mpl{M_{\rm Pl}}
\def\ms{M_{\rm s}}
\def\ls{l_{\rm s}}
\def\l{\lambda}
\def\m{\mu}
\def\n{\nu}
\def\a{\alpha}
\def\b{\beta}
\def\gs{g_{\rm s}}
\def\d{\partial}
\def\co{{\cal O}}
\def\sp{\;\;\;,\;\;\;}
\def\r{\rho}
\def\dr{\dot r}

\def\e{\epsilon}
\newcommand{\NPB}[3]{\emph{ Nucl.~Phys.} \textbf{B#1} (#2) #3}   
\newcommand{\PLB}[3]{\emph{ Phys.~Lett.} \textbf{B#1} (#2) #3}   
\newcommand{\ttbs}{\char'134}        
\newcommand\fverb{\setbox\pippobox=\hbox\bgroup\verb}
\newcommand\fverbdo{\egroup\medskip\noindent%
                        \fbox{\unhbox\pippobox}\ }
\newcommand\fverbit{\egroup\item[\fbox{\unhbox\pippobox}]}
\newbox\pippobox
\def\tr{\tilde\rho}
\def\lb{w}
\def\bbox{\nabla^2}
\def\mt{{\tilde m}}
\def\rct{{\tilde r}_c}

\def \lta {\mathrel{\vcenter
     {\hbox{$<$}\nointerlineskip\hbox{$\sim$}}}}
\def \gta {\mathrel{\vcenter
     {\hbox{$>$}\nointerlineskip\hbox{$\sim$}}}}

\noindent
\begin{flushright}

\end{flushright} 
\vspace{1cm}
\begin{center}
{ \Large \bf Nonlinear Matter Spectra 
in Growing Neutrino Quintessence \\}
\vspace{0.5cm}
{N. Brouzakis$^{(1)}$, 
V. Pettorino$^{(2)}$,
N. Tetradis$^{(1)}$, 
C. Wetterich$^{(3)}$ } 
\end{center}
\vspace{0.6cm}
(1) {\it Department of Physics, University of Athens, 
University Campus, Zographou 157 84, Greece
\\
(2)
{\it SISSA, Via  Bonomea 265, 34136 Trieste, Italy}
\\
(3) {\it Institut f\"ur Theoretische Physik, Universit\"at Heidelberg, Philosophenweg 16,
Heidelberg 69120, Germany}
} 
\vspace{1cm}
\abstract{
We investigate the nonlinear power spectra of density perturbations and acoustic oscillations in growing neutrino
quintessence. In this scenario, the neutrino mass has a strong dependence on the quintessence 
field. The induced coupling stops the evolution of the field when the neutrinos become nonrelativistic, 
and triggers the transition to the accelerating phase of the cosmological expansion.
For the calculation of the nonlinear spectra we employ the time renormalization group, which 
resums subsets of diagrams of arbitrarily high order in cosmological perturbation theory. 
At redshifts around five, the neutrino fluctuations are still linear and acoustic oscillations are present in the neutrino
power spectrum, induced by the  acoustic oscillations in the baryonic and dark-matter sectors.
The neutrino perturbations become nonlinear at redshifts 
around three. The mode coupling generated by the nonlinearities erases the oscillations in the neutrino spectrum
at some redshift above two. There is a potential danger that at later times the 
influence of the gravitational potentials induced by the neutrino inhomogeneities 
could erase the 
oscillations from the baryonic and dark-matter spectra, making the scenario incompatible with 
observations. For the scenario to be viable, the neutrino-induced gravitational potentials in the range of 
baryon acoustic oscillations should not grow to average values much larger than $10^{-4}$. 
The magnitude of the expected potentials is still not known reliably, as the  process of structure formation 
is poorly understood in growing neutrino quintessence. The time renormalization group cannot describe the
effects of nonlinear clustering. Alternative methods, such as hydrodynamic simulations, must be empoloyed
for the calculation of the spectra at low redshifts. 
\\~\\
\vspace{2cm}
PACS numbers: 95.36.+x, 13.15.+g, 98.80.Es
}

\newpage

\section{Introduction}
\setcounter{equation}{0}

A popular extension of the quintessence scenario \cite{quintessence} assumes the presence of a coupling
between the dark-energy and dark-matter sectors \cite{dmde}. This assumption provides an extended framework
in which one may hope  to address the coincidence problem, i.e. the reason behind the comparable present 
contributions from dark matter and dark energy to the total energy  density. In a variation of this scenario, characterized
as growing neutrino quintessence, 
the interaction with dark energy is shifted from the dark matter to the cosmological neutrino sector \cite{growing}.
The neutrino-dark energy coupling $\beta$ can be large, and generate a force substantially stronger than
the standard gravitational interaction \cite{mvn}.
As a result, even if the neutrinos contribute only a small fraction to the total energy density, they may have 
a significant effect on the cosmological evolution \cite{growing}. Their effect becomes important when the
neutrino mass stops being negligible and the neutrinos become nonrelativistic. This happens at 
a redshift $z_{nr}\sim 5-10$, with the exact value depending on the particular model. At lower redshifts, the 
presence of the neutrinos forces the quintessence field to stop evolving, so that its potential acts as
an effective cosmological constant, whose value is related to the present neutrino mass. 

The presence of an additional force in the neutrino sector, which is stronger than gravity by a factor
$\sim \beta^2$, has profound implications for the evolution of cosmological neutrino perturbations. 
As soon as neutrinos become nonrelativistic ($z < z_{nr}$), the perturbations in their energy density start
to grow, with a characteristic timescale that is shorter by a factor $\sim \beta^{-2}$ relative 
to the one characterizing standard gravitational collapse. Static solutions of Einstein's equations 
are known, which describe neutrino lumps held together by the force mediated by the quintessence field. 
Such configurations may be the endpoint of the collapse process \cite{lumps}. The analysis of the growth of neutrino
perturbations beyond linear level is complicated and the final stages are not well understood \cite{collapse}.
The gravitational potential induced by large-scale structures in the neutrino sector, which can have a size
of 100 Mpc or more, can affect the propagation of photons and influence the cosmic microwave background (CMB) 
through the
integrated Sachs-Wolfe (ISW) effect. An analysis based on linear perturbation theory leads to the 
conclusion that models with a strong neutrino-dark energy coupling are excluded \cite{cmb}.
However, an extrapolation of linear growth is strongly misleading, as it would imply completely unrealistic 
neutrino-induced cosmological gravitational potentials \cite{cmb2}. In particular, 
the effects of backreaction and virialization are not accounted for in a linear treatment.
If these are sufficiently strong, the neutrino structures may not be as dense as the extrapolation
of the linear analysis would indicate. The resulting gravitational potentials may be sufficiently small
so as not to affect the CMB significantly \cite{cmb2}. As a general rule of thumb, potentials larger than
$\sim 10^{-5}$ at length scales of 100 Mpc or more give too strong an effect on the CMB. 

In this work we study the power spectra of dark-matter and neutrino perturbations in the scenario of growing neutrino quintessence.
The growth of perturbations can be used in order to
constrain the scenario through comparison with the observed large-scale structure.
The most promiment feature of the baryonic and dark-matter spectrum is a series of peaks and valleys, characterized as 
baryon acoustic oscillations (BAO). They originate in the period of recombination, and correspond to
sound waves in the relativistic plasma of that epoch. 
The chacteristic length scale of BAO is around
100 Mpc. Even in the standard $\Lambda$CDM scenario,
the exact form of the dark-matter power spectrum at such scales is not easy to compute precisely, because of 
the failure of linear perturbation theory to describe reliably the growth of the corresponding  
fluctuations under gravitational collapse.   
At length scales below about 10 Mpc, the evolution is highly
non-linear, so that only numerical N-body simulations can capture the dynamics of the formation 
of galaxies and clusters of galaxies. However, fluctuations with length scales of 
around 100 Mpc fall within the mildly non-linear regime, for which analytical methods
have been developed. We focus on scales in the range 50--200 Mpc, within which BAO are visible. 
In growing neutrino quintessence the neutrino power spectrum displays a much faster growth and overtakes
the dark-matter spectrum at redshifts below $z\sim 4$. The nonlinear corrections become very large, so that analytical
methods become unreliable even in the BAO range at redshifts near $z=0$. Our aim
is to explore the range of validity of the analytical methods and investigate the form of the spectra within this range.

The various analytical methods \cite{CrSc1}--\cite{matsubara} that have been developed 
in order to go beyond linear perturbation theory essentially amount to 
resummations of subsets of perturbative diagrams of arbitrarily high order, in a way analogous to the 
renormalization group (RG).
We follow the approach of  \cite{Max2}, named time-RG or TRG, which 
uses time as the flow parameter that describes the evolution of physical quantities,
such as the spectra. The method 
has been applied to ${\rm \Lambda CDM}$ and quintessence cosmologies \cite{Max2}, allowing for 
a possible coupling of dark energy to dark matter \cite{coupled}, or a variable equation of state \cite{wz}. It has also been
used for the study of models with neutrinos of constant mass \cite{lesgourgues}.  A comparative analysis of 
several analytical methods, using N-body simulations as 
a reference, has been carried out in ref. \cite{carlson} for $\Lambda$CDM cosmology. The study  
demonstrates that TRG remains accurate at the 1-2\% level over the whole BAO range  at all redshifts. 

It must be emphasized that all the methods that amount in resummations of perturbative diagrams have
a limited range of validity. 
As they are based on the single-stream approximation, they are
applicable only when multistreaming, i.e. non vanishing velocity
dispersion, can be neglected. For $\Lambda$CDM near $z=0$, this regime includes the BAO range, in which 
the resummation methods provide a significant improvement to linear perturbation theory. 
However, at smaller length scales, in which the process of halo formation is important, these methods are not
applicable. (An attempt to extend the range of applicability of TRG is described in ref. \cite{halores}.)
The physical processes which dominate the halo dynamics can be capured reliably only through
hydronamic simulations, which are, however, more time consuming.

In the following section we summarize the formalism we use. In section 3 we present the results of the numerical 
integration of the evolution equations for the spectra. Finally, in section 4 we give our conclusions.

\section{Neutrinos coupled to dark energy}
\setcounter{equation}{0}

\subsection{Evolution equations for the perturbations}

We assume that the energy density of the Universe 
receives significant contributions from three components: 
a) nonrelativistic matter, in which we group standard baryonic matter (BM) and cold dark matter (CDM); 
b) massive neutrinos, whose equation of state varies during the cosmological evolution;
c) a slowly varying, classical scalar field $\phi$, named cosmon \cite{quintessence}, whose contribution to the energy
density is characterized as dark energy (DE).
We also allow for a direct coupling 
between the neutrinos and the cosmon field. In particular, we assume that the 
mass $m_\nu$ of the neutrinos has a dependence on $\phi$, and define the coupling as
$\beta(\phi)=-{d\ln m_\nu(\phi)}/{d\phi}$. In the present paper we consider a model with constant $\beta$.
The equation of motion of the cosmon field takes the form
\be
\frac{1}{\sqrt{-g}}\frac{\partial}{\partial x^\mu}
\left(\sqrt{-g}\,\,g^{\mu\nu}\frac{\partial \phi}{\partial x^\nu}
\right)
=-\frac{dU}{d\phi}+{\beta(\phi)}\,\, \left( T_{\nu} \right)^\mu_{~\mu}.
\label{eomphia} \ee
At early times, when the neutrinos are relativistic and their energy-momentum tensor is traceless, they 
are effectively decoupled from the cosmon field. Only when the neutrinos become nonrelativistic the cosmon-neutrino coupling
$\beta$ becomes effective, leading to
energy exchange between the neutrino and dark-energy sectors.
We normalize all dimensionful quantities, such as the
cosmon field, with respect to the reduced Planck mass $M=(8\pi G)^{-1/2}$. 
This is equivalent to setting $M=1$. 

We are interested in the more recent stages of the cosmological evolution (redshifts $z\lta 10$). 
In the scenario we consider, the neutrinos become nonrelativistic at such redshifts.  
In the remaining of the section, in which we develop the formalism for the treatment of the nonlinear corrections
to the power spectra, we assume that the neutrino pressure vanishes. 
We approximate the metric as 
\be
ds^2=a^2(\tau)\left[
\left(1+2\Phi(\tau,\vec{x}) \right)d\tau^2
-\left(1-2\Phi(\tau,\vec{x}) \right) d\vec{x}\, d\vec{x} \right].
\label{metric} \ee
We assume that the Newtonian potential $\Phi$ is weak: $\Phi \ll 1$.
Also we decompose the cosmon field $\phi$ as
$\phi(\tau,\vec{x})=\bar{\phi}(\tau)+\dphi(\tau,\vec{x})$,
with $\dphi \ll 1$. In general,
$\bar{\phi}={\cal O}(1)$ in units of $M$.
Finally, we decompose the density as
$\rho(\tau,\vec{x})=\bar{\rho}(\tau)+\drho(\tau,\vec{x})$,
while we neglect the pressure.
A self-consistent expansion scheme can be obtained if we assume the  
hierarchy of
scales: $\Phi, \dphi \ll |\dvvec | \ll \drho/\bar{\rho}  \lesssim 1$. 
Such a hierarchy is predicted by the linear analysis for subhorizon perturbations with momenta $k\gg \Hc=\dot{a}/a$.
For subhorizon perturbations, it is consistent to make
the additional assumption that the spatial derivatives of $\Phi, \dphi$
dominate over the time derivatives. 
The predictions of the linear analysis allow us to make a more quantitative
statement. We assume that a spatial derivative acting on $\Phi$, $\dphi$ or 
$\dvvec$ increases the position of that quantity in the hierarchy by one 
level. In this sense $\vec{\nabla} \Phi$ is comparable to $\Hc \dvvec$, while
$\nabla^2 \Phi$ is comparable to $\Hc^2 \delta\rho/\bar{\rho}$. 

With the above assumptions, one can derive the equations 
that describe the evolution of the Universe. The details of the calculation are given in \cite{coupled}.
The evolution of the homogeneous background is described by 
\begin{subequations}
\begin{align}
\mathcal{H}^{2}=&\dfrac{1}{3}\big[a^{2}\sum_{i=1,2}\bar{\rho}_{i}
+\frac{1}{2}{\dot{\bar{\phi}}^{2}}+a^{2} U(\bar{\phi})\big]
\equiv\dfrac{1}{3}{a^{2}\rho_{tot}}\\
\dot{\bar{\rho}}_{i}+3\mathcal{H}\bar{\rho}_{i}
=&-\beta_{i}\,\dot{\bar{\phi}}\bar{\rho}_{i}\\
\ddot{\bar{\phi}}+2\mathcal{H}\dot{\bar{\phi}}=&-{a^{2}}
\left(\dfrac{dU}{d\phi}(\bar{\phi})-\sum_{i=1,2}\beta_{i}\,\bar{\rho}_{i}\right),
\label{25c} \end{align}
\end{subequations}
where we have defined $\rho_{tot}\equiv\sum_{i}\bar{\rho}_{i}
+{\dot{\bar{\phi}}^{2}}/{(2a^{2})}+U(\bar{\phi})$. The values $i=1,2$  correspond to neutrinos and 
CDM+BM, respectively. For the neutrinos we consider a constant coupling
$\beta_1=\beta$,  while we assume that the CDM+ BM sector does not couple to the cosmon field:
$\beta_2=0$.
 
We describe the perturbations in terms of the cosmon perturbation $\delta\phi$,
the Newtonian potential $\Phi$, the density perturbations $\delta\rho_i$ and the 
velocity fields $v_i$.
We have two Poisson equations 
\begin{subequations}
 \begin{align}
  \nabla^{2}\delta\phi=&-{a^{2}\sum_{i}\beta_{i}\delta\rho_{i}}
\equiv -3\sum_{i}\beta_{i}\mathcal{H}^{2}\Omega_{i}\delta_{i} \label{poiss1} \\
\nabla^{2}\Phi=&\frac{1}{2}a^{2}{\sum_{i}\delta\rho_{i}}
\equiv\dfrac{3}{2}\mathcal{H}^{2}\sum_{i}\Omega_{i}\delta_{i},
\label{poiss2} \end{align}
\end{subequations}
with 
$\Omega_{i}(\tau)
\equiv{\bar{\rho}_{i}\,a^{2}}/{(3\mathcal{H}^{2})}$,
and the continuity and Euler equations
\begin{subequations}
 \begin{align}
{\delta\dot{\rho}}_{i}+3\mathcal{H}\delta\rho_{i}+\vec{\nabla}
[(\bar{\rho}_{i}+\delta\rho_{i})\delta{\vec{v}_i}]=&-\beta_{i}\dot{\bar{\phi}}
\delta\rho_{i}\label{a}\\
 \delta \dot{\vec{v}}_{i}+(\mathcal{H}-\beta_{i}\dot{\bar{\phi}})
\delta \vec{v}_{i}+(\delta{\vec{v}}_{i}\cdot\vec{\nabla})\delta{\vec{v}}_{i}
=&-\vec{\nabla}\Phi+\beta_{i}\vec{\nabla}\delta\phi\label{b}. 
 \end{align}
\end{subequations}

The inspection of eqs. (\ref{poiss1})-(\ref{b}) reveals a potential shortcoming of the assumed hierarchy for large $\beta$.
From the Euler equation one concludes that the scalar quantity comparable to the gravitational
potential is actually $\beta \delta\phi$.
From eqs. (\ref{poiss1}), (\ref{poiss2}) we can see that $\beta \dphi$ is typically larger than $\Phi$ by a 
factor $\sim \beta^2$. This limits the range of validity of a hierarchy in which $\beta\dphi$ and $\Phi$ are 
treated on equal footing. Furthermore, eqs. (\ref{a}), (\ref{b}) implicitly neglect effects arising from a
difference between the local neutrino mass, as given by the local value of the cosmon field,
and the average cosmological neutrino mass, as given by the background value $\bar{\phi}$. Including higher
orders in $\beta \dphi$ wold increase considerably the algebraic complexity of the evolution equations.  

\subsection{Evolution equations for the power spectra}
The evolution equations are expressed in their most 
useful form in terms of the density contrasts 
$\delta_{i}\equiv{\delta\rho_{i}}/{\bar{\rho}_{i}}\lesssim 1$
and the divergence of the velocity field
$\theta_{i}(\textbf{k}, \tau)\equiv\vec{\nabla}\cdot\vec{\delta v_{i}}(\textbf{k}, \tau)$.
For the Fourier transformed quantities, 
we obtain from eq. (\ref{a})
\begin{equation}\label{delta}
\dot{\delta_{i}}(\textbf{k}, \tau)+\theta_{i}(\textbf{k}, \tau)
+\int d^{3}\textbf{k}_{1} \,d^{3}\textbf{k}_{2}\,
\delta_{D}(\textbf{k}-\textbf{k}_{1}-\textbf{k}_{2})\,
\alpht(\textbf{k}_{1}, \textbf{k}_{2})\,
\theta_{i}(\textbf{k}_{1}, \tau)\,
\delta_{i}(\textbf{k}_{2}, \tau)
=0,
\end{equation}
where 
\begin{equation}
\alpht(\textbf{k}_{1}, \textbf{k}_{2})=\dfrac{\textbf{k}_{1}\cdot(\textbf{k}_{1}
+\textbf{k}_{2})}{k_{1}^{2}}. 
\end{equation}
Eqs. (\ref{b}), (\ref{poiss1}), (\ref{poiss2}) give
\begin{equation}\label{theta}
\begin{split}
\dot{\theta}_{i}(\textbf{k}, \tau)+
&(\mathcal{H}-\beta_{i}\dot{\bar{\phi}})\theta_{i}(\textbf{k}, \tau)
+\dfrac{3\mathcal{H}^{2}\sum_{j}\Omega_{j}(2\beta_i\beta_{j}+1)\delta_{j}
(\textbf{k}, \tau)}{2}\\
+&\int d^{3}\textbf{k}_{1}\, d^{3}\textbf{k}_{2}\,
\delta_{D}(\textbf{k}-\textbf{k}_{1}-\textbf{k}_{2})\,
{\bett}(\textbf{k}_{1}, \textbf{k}_{2})\,
\theta_{i}(\textbf{k}_{1}, \tau)\,
\theta_{i}(\textbf{k}_{2}, \tau)=0,
\end{split}
\end{equation}
where 
\begin{equation}
 \bett(\textbf{k}_{1}, \textbf{k}_{2})=\dfrac{(\textbf{k}_{1}+\textbf{k}_{2})^{2}
\textbf{k}_{1}\cdot\textbf{k}_{2}}{2 k_{1}^{2}k_{2}^{2}}.
\end{equation}

We define the quadruplet
\begin{equation}
 \left(
\begin{array}{c}
\varphi_{1}(\textbf{k}, \eta)\\ \\ \varphi_{2}(\textbf{k}, \eta)\\ \\ 
\varphi_{3}(\textbf{k}, \eta)\\ \\ \varphi_{4}(\textbf{k}, \eta)\end{array}
\right)
=e^{-\eta}\left(
\begin{array}{c}
\delta_{\nu}(\textbf{k}, \eta)\\ \\-\dfrac{\theta_{\nu}(\textbf{k}, \eta)}{\mathcal{H}}\\ \\ 
\delta_{m}(\textbf{k}, \eta)\\ \\-\dfrac{\theta_{m}(\textbf{k}, \eta)}{\mathcal{H}}
\end{array}
\right),
\label{quadruplet}
\end{equation}
where $\eta=\ln[ a(\tau)/a_i]$, with $a_i$ the scale factor at some convenient time, at which we define the initial conditions. 
(We normalize the scale factor so that $a_0=1$ today.)
The index $\nu$ refers to neutrinos and $m$ to the CDM+BM sector.   
This allows us to bring 
eqs. (\ref{delta}), (\ref{theta}) 
in the form \cite{CrSc1,Max1,Max2} 
\begin{equation}\label{arghh}
\partial_{\eta}\varphi_{a}(\textbf{k}, \eta)+\Omega_{ab}\varphi_{b}(\textbf{k}, \eta)
=e^{\eta}\gamma_{abc}(\textbf{k}, -\textbf{k}_{1}, -\textbf{k}_{2})
\varphi_{b}(\textbf{k}_{1}, \eta)\varphi_{c}(\textbf{k}_{2}, \eta).
\end{equation}
The indices $a,b,c$ take values $1,\ldots, 4$.
The values 1, 2
characterize neutrino density and velocity perturbations, while 3, 4 refer
to CDM+BM quantities.  
Repeated momenta are integrated over, while repeated indices are summed over. 
The functions $\gamma$, that determine effective vertices, 
are analogous to those employed in \cite{Max1,Max2}.
The non-zero components are
\begin{equation}
\begin{split}
\gamma_{121}(\textbf{k},\textbf{k}_{1}, \textbf{k}_{2})&
=\dfrac{\alpht(\textbf{k}_{1}, \textbf{k}_{2})}{2}
\delta_{D}(\textbf{k}+\textbf{k}_{1}+\textbf{k}_{2})
=\gamma_{112}(\textbf{k}, \textbf{k}_{2}, \textbf{k}_{1})\\
\gamma_{222}(\textbf{k},\textbf{k}_{1}, \textbf{k}_{2})
&=\bett(\textbf{k}_{1}, \textbf{k}_{2})\ 
\delta_{D}(\textbf{k}+\textbf{k}_{1}+\textbf{k}_{2})\\
\gamma_{343}(\textbf{k},\textbf{k}_{3}, \textbf{k}_{4})
&=\dfrac{\alpht(\textbf{k}_{3}, \textbf{k}_{4})}{2}
\delta_{D}(\textbf{k}+\textbf{k}_{3}+\textbf{k}_{4})
=\gamma_{334}(\textbf{k}, \textbf{k}_{4}, \textbf{k}_{3})\\
\gamma_{444}(\textbf{k},\textbf{k}_{3}, \textbf{k}_{4})
&=\bett(\textbf{k}_{3}, \textbf{k}_{4})\ 
\delta_{D}(\textbf{k}+\textbf{k}_{3}+\textbf{k}_{4}).
\end{split}
\label{vertex}
\end{equation}
The $\Omega$-matrix is
\be
 \Omega(\eta)=\left(
\begin{array}{cccc}
 1 & -1 & 0 & 0
\\ \\
-\dfrac{3}{2}\Omega_{\nu}(2\beta^{2}+1) & 2-\beta\bar{\phi}'+\dfrac{\mathcal{H}'}{\mathcal{H}} & -\dfrac{3}{2}\Omega_{m} & 0
\\ \\ 
0 & 0 & 1 & -1
\\ \\
-\dfrac{3}{2}\Omega_{\nu} & 0 & -\dfrac{3}{2}\Omega_{m} & 2+\dfrac{\mathcal{H}'}{\mathcal{H}}
\end{array}
\right),
\label{omegafour}
\ee
where a prime denotes a derivative with respect to $\eta$.

The next step is to derive evolution equations for the power spectra.  
The spectra and bispectra are defined as
\begin{equation}
\begin{split}
 \langle\varphi_{a}(\textbf{k}, \eta)\varphi_{b}(\textbf{q}, \eta)\rangle
\equiv&
\delta_{D}(\textbf{k}+\textbf{q}) P_{ab}(\textbf{k}, \eta)\\
\langle\varphi_{a}(\textbf{k}, \eta)\varphi_{b}(\textbf{q}, \eta)\varphi_{c}(\textbf{p}, \eta)\rangle
\equiv&
\delta_{D}(\textbf{k}+\textbf{q}+\textbf{p}) B_{abc}(\textbf{k}, \textbf{q},\textbf{p},\eta).
\end{split}
\label{spectra} \end{equation}
The essential approximation that we have to make in order to obtain a closed 
system of equations is a truncation of the four-point function which appears in the evolution equation for the 
trispectrum. In this way we obtain \cite{Max2,coupled}
\begin{eqnarray}
\partial_{\eta}P_{ab}(\textbf{k}, \eta)&=&-\Omega_{ac}P_{cb}(\textbf{k}, 
\eta)-\Omega_{bc}P_{ac}(\textbf{k}, \eta)
\nonumber \\
&&+e^{\eta}\int d^{3}q \big[\gamma_{acd}(\textbf{k},-\textbf{q}, \textbf{q}-\textbf{k})
B_{bcd}(\textbf{k},-\textbf{q}, \textbf{q}-\textbf{k})
\nonumber \\
&&+\gamma_{bcd}(\textbf{k},
-\textbf{q}, \textbf{q}-\textbf{k})B_{acd}(\textbf{k},-\textbf{q}, \textbf{q}
-\textbf{k})\big],
\label{spectev1}\\
 \partial_{\eta}B_{abc}(\textbf{k},-\textbf{q}, \textbf{q}-\textbf{k})&=&
-\Omega_{ad}B_{dbc}(\textbf{k},-\textbf{q}, \textbf{q}-\textbf{k})-\Omega_{bd}B_{adc}
(\textbf{k},-\textbf{q}, \textbf{q}-\textbf{k})-\Omega_{cd}B_{abd}(\textbf{k},
-\textbf{q}, \textbf{q}-\textbf{k})
\nonumber \\
&&+2e^{\eta} \big[\gamma_{ade}(\textbf{k},-\textbf{q}, \textbf{q}-\textbf{k})
P_{db}(\textbf{q}, \eta)P_{ec}(\textbf{k}-\textbf{q}, \eta)
\nonumber \\
&&+\gamma_{bde}(-\textbf{q},\textbf{q}-\textbf{k}, \textbf{k})P_{dc}(\textbf{k}
-\textbf{q}, \eta)P_{ea}(\textbf{k}, \eta)
\nonumber \\
&&+\gamma_{cde}(\textbf{q}-\textbf{k}, 
\textbf{k}, -\textbf{q})P_{da}(\textbf{k}, \eta)P_{eb}(\textbf{q}, \eta)\big].
\label{spectev2}
\end{eqnarray}

\section{Numerical analysis}
\setcounter{equation}{0}

\subsection{Approximations}

The presence of two massive species (neutrinos and CDM+BM) complicates the structure of the equations compared to the case where they are 
treated as a single fluid,  discussed in \cite{Max2}. The full system of eqs.~(\ref{spectev1}), (\ref{spectev2}) contains 74 equations, namely, 10 for the power spectra and 64 for the bispectra, compared to the 11 equations of the single-matter case \cite{Max2}. An accurate 
calculation also requires the discretization of the $k$-space with at least 500 points. Taking into account that the bispectra depend 
on three  external momenta, it is apparent that the necessary computing power is significant. 
In order to make the numerical integration of the evolution equations feasible, we have to make additional approximations.

As will be apparent in the following, the method we are employing remains valid only within the mildly nonlinear regime.
Strong nonlinearities and backreaction effects, as well as the process of virialization, are not accounted for in our treatment.
In the scenario we are considering, the strongest nonlinearities appear in the neutrino sector, while the
CDM+BM spectrum remains within the linear regime. It is, therefore, a good approximation to neglect the 
nonlinear terms in the evolution equations for the CDM+BM spectrum.
In practice, this means that we may set the vertices $\gamma_{334}$,  $\gamma_{343}$,  $\gamma_{444}$, defined in
eq. (\ref{vertex}), equal to zero. 

We derive an approximate solution of eqs.~(\ref{spectev1}), (\ref{spectev2}) in the following way:
\begin{itemize}
\item
At a first stage, we integrate the full system of equations setting all the vertices $\gamma_{abc}$ equal to zero. This reproduces the 
linear spectra. 
\item
Subsequently, we integrate the 11 equations for $P_{ab}$, $B_{abc}$ with $a,b=1,2$. Notice that these involve only the vertices
 $\gamma_{112}$,  $\gamma_{121}$,  $\gamma_{222}$. All spectra and bispectra appearing in these equations with an index
3 or 4, because of the nonzero entries  $\Omega_{23}$ and $\Omega_{41}$ entries of eq.~(\ref{omegafour}), 
are approximated by their linear solutions derived at the previous stage.
\end{itemize}
The above procedure can reproduce the modifications in the neutrino spectrum induced by mode coupling at the nonlinear
level. Our main result will be the effect of this process on the BAO in the spectrum. Our analysis is performed at redshifts 
above $z\sim 2$, at which the only significant nonlinearities appear in the neutrino sector. 

\begin{figure}[t]
\begin{center}
\includegraphics[width=0.8\textwidth]{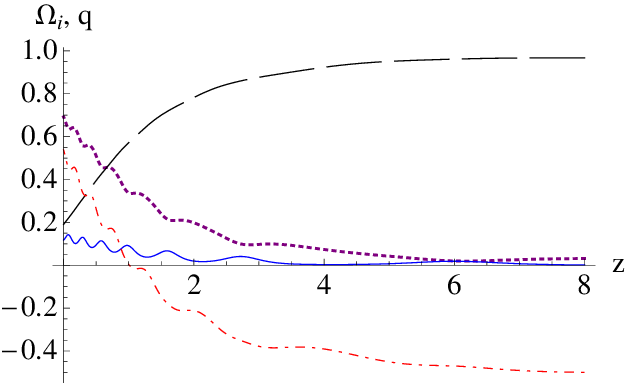}
\end{center}
\caption {The fractional energy density in neutrinos (solid), CDM+BM (dashed) and dark energy (dotted). The acceleration parameter
$q=a\ddot{a}/\dot{a}^2$ (dot-dashed) is also depicted.} 
\label{background}
\end{figure}

\begin{figure}[t]
\begin{center}
\includegraphics[width=0.8\textwidth]{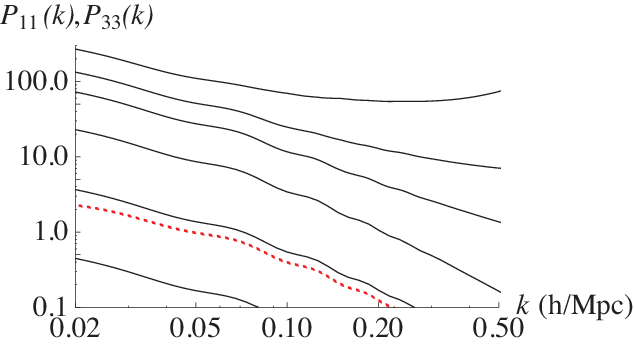}
\end{center}
\caption { The neutrino density power spectrum
$P_{11}(k,\eta)$ (solid lines) and the CDM+BM density spectrum $P_{33}(k,\eta)$ (dotted lines) at 
redshifts $z=4.70$, 4.08, 3.04, 2.77, 2.69, 2.60 (starting from below). } 
\label{spectraa}
\end{figure}

\begin{figure}[t]
\begin{center}
\includegraphics[width=0.8\textwidth]{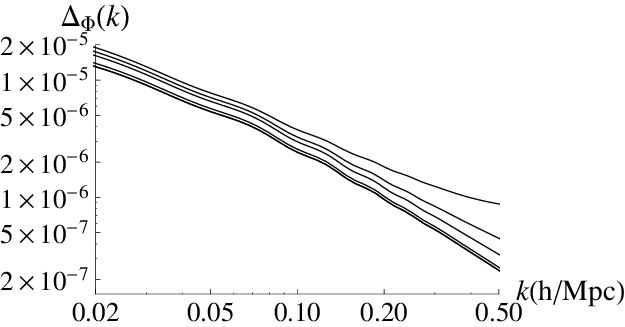}
\end{center}
\caption {The typical size of the gravitational potential as a function of the scale $k$ at 
redshifts $z=4.70$, 4.08, 3.04, 2.77, 2.69, 2.60 (starting from below). } 
\label{potential}
\end{figure}

\subsection{Results}

We consider a model in which the cosmon field has a potential $V(\phi)=M^4 \exp (-\alpha \phi)$, with $\alpha=10$. 
As we mentioned earlier, we normalize dimensionful quantities with respect to the Planck mass $M=1$.
The coupling between the neutrinos and the cosmon field is taken to be constant. This is equivalent
to assuming that the neutrino mass is $m_\nu(\phi)=\tilde{m}_\nu \exp (-\beta \phi)$. 
In our calculation we consider three degenerate neutrinos and use $\beta=-52$.
The constant $\tilde{m}_\nu$ and the present value of the background field $\phi$ are chosen such that the present neutrino mass is
$m_{\nu 0}=2.1$ eV. The evolution of the cosmological background for $z<z_{nr}=8$ is shown in fig. \ref{background}. We have 
approximated the neutrinos as nonrelativistic during this period. We depict the fractional energy density in neutrinos (solid), 
CDM+BM (dashed) and dark energy (dotted). We also plot the acceleration parameter $q=a\ddot{a}/\dot{a}^2$
(dot-dashed). The cosmological expansion becomes accelerating at a redshift $z \simeq 1$.
The present fractional contributions of neutrinos, CDM+BM and 
the cosmon field to the total energy density are $\Omega_\nu=0.13$, $\Omega_m=0.23$, $\Omega_\phi=0.64$, respectively. 
The spatial 
curvature is assumed to vanish, while the current expanstion rate is $H_0=72 $ km s$^{-1}$ Mpc$^{-1}$.

We start the integration of the evolution equations for the spectra at a redshift $z_{nr}=8$. The neutrinos are 
already nonrelativistic at this time in our model. The initial conditions at $z_{nr}$ are obtained through the implementation of the 
background and linear-perturbation equations in the code CAMB \cite{camb}, generalized for the case of 
mass-varying neutrinos. We assume that the primordial spectrum is scale invariant with spectral index close to 1.
In the background evolution, we take into account the transition of neutrinos from being relativistic at high redshifts to nonrelativistic
near $z_{nr}$. We point out that the neutrinos are essentially decoupled from the cosmon field during the time that they
are relativistic (see eq. (\ref{eomphia})).
We perform a careful analysis of the evolution of the neutrino perturbations by solving the 
Boltzmann equation. In this way we obtain the neutrino and matter spectra at $z_{nr}$. 
The mixed spectra are obtained 
as the geometrical averages of the pure ones, i.e.
 \begin{equation}\label{approxspectra}
P_{12}(k,\eta_{nr})\simeq\sqrt{P_{11}(k,\eta_{nr})P_{22}(k,\eta_{nr})},
~~~~~~
P_{13}(k,\eta_{nr})\simeq\sqrt{P_{11}(k,\eta_{nr})P_{33}(k,\eta_{nr})},
\end{equation}
etc, consistently with the expectation in the linear regime.

The evolution of the spectra at redshifts below $z_{nr}$ is linear to a good approximation down to $z\sim 3$. 
Around this time the neutrino spectrum has grown sufficiently for the nonlinear corrections to affect the 
evolution. This is demonstrated in fig. \ref{spectraa}, where we depict the neutrino density power spectrum
$P_{11}(k,\eta)$ (solid lines) and the CDM+BM density spectrum $P_{33}(k,\eta)$ (dotted lines) at various redshifts.
The solid lines, starting from below, correspond to $z=4.70$, 4.08, 3.04, 2.77, 2.69, 2.60. The dotted lines correspond
to the same redshifts, but  they almost coincide. We point out that the actual density spectra are obtained from 
those depicted in fig. \ref{spectraa} by multiplication with a factor $\exp(2\eta)=(a(z)/a(z_i))^2=81(1+z)^{-2}$, where we have
taken $z_i=z_{nr}=8$ in our calculation. This factor arises because of the definitions (\ref{quadruplet}), (\ref{spectra}), while
$\eta=\ln[ a(\tau)/a_i]$, with $a_i$ the scale factor at some convenient time, at which we define the initial conditions. 
In our calculation $a_i=1/9$.  The spectra are given in units of $({\rm Mpc}/h)^3$, consistently with their definition (\ref{spectra}).
They are also smaller by a factor $(2\pi)^3$ compared to the definition in CAMB.

At high redshifts the neutrino density spectrum is very suppressed because of free-streaming, as the neutrinos are relativistic over
the entire depicted momentum range. 
It is apparent from fig. \ref{spectraa} that, as soon as the neutrinos become nonrelativistic, their 
spectrum grows very fast because of the additional
attractive force generated by the neutrino-dark energy coupling. This force is $2\beta^2\simeq 5400$ times stronger than gravity. 
The neutrino density spectrum at $z=4.70$, 4.08, 3.04 is within the linear regime, while the nonlinearities become 
apparent at $z=$2.77, 2.69, 2.60: starting from  large $k$, the neutrino spectrum grows faster than the linear analysis
would predict. The matter spectrum retains its shape, as its evolution is described by linear theory to a very good
approximation for the whole range of redshifts depicted in fig. \ref{spectraa}.

The most interesting feature of the spectra in fig. \ref{spectraa} is their oscillatory behavior. The BAO in the matter sector
originate in the period of recombination, and correspond to
sound waves in the relativistic plasma of that epoch. The corresponding oscillations in the neutrino spectrum develop
as soon as the neutrinos become nonrelativistic and start falling in the gravitational potentials generated by the
CDM+BM inhomogeneities. The corresponding process takes place within the linear regime and is fast, as the 
CDM+BM power spectrum exceeds that of neutrinos by several orders of magnitude. 
The amplitude of oscillations in the neutrino spectrum starts diminishing as soon as the nonlinear corrections become important.
Again the transition is fast, and the oscillations are hardly visible at $z=2.60$. This phenomenon has its origin in
the mode coupling induced by the nonlinearities. It is observed within the $\Lambda$CDM model as well: near $z=0$ the 
higher peaks in the matter spectrum disappear and only the first two or three survive when the nonlinear corrections
are taken into account. In the model at hand, the nonlinearities in the neutrino sector are very strong, because of the
rapid growth of the spectrum. The oscillations disappear at a redshift above $z\sim 2$.

A crucial question is whether the disappearance of oscillations from the neutrino sector induces their elimination from the 
CDM+BM sector as well. We address this issue in the following section. An important quantity in the context of
this discussion is the 
gravitational potential induced by the inhomogeneities. We can have an estimate of its magnitude at various 
scales by starting from the Poisson equation (\ref{poiss2}) for the gravitational potential and defining an associated power 
spectrum as 
\begin{equation}
\Delta^2_\Phi(k,\eta)=4 \pi k^3 P_\Phi(k,\eta)=4\pi k^3 e^{2\eta} \left( \frac{3}{2} \frac{\Hc^2}{k^2} \right)^2
\left( \Omega^2_\nu P_{11}+2 \Omega_\nu\Omega_m P_{12} +\Omega_m^2 P_{22} \right).
\label{dphi} \end{equation}
In fig. \ref{potential} we plot $\Delta_\Phi$ as a function of $k$ for redshifts $z=4.70$, 4.08, 3.04, 2.77, 2.69, 2.60.
At $z=4.70$ the potential is mainly generated by the CDM+BM inhomogeneities, while for $z=2.60$ the main
contribution at large $k$ comes from the neutrino inhomogeneities. This is apparent in the change of shape of the
function $\Delta_\Phi(k)$. Already at this stage the potential takes values in the region
$[10^{-6}-10^{-5}]$ in the BAO range. At $z\lta 2.6$ the neutrinos dominate the gravitational
potential, so that its magnitude depends on the growth of inhomogeneities in this sector. 
We point out that $k$ is the comoving momentum, so that the physical length scale 
is related to $k/a(\eta)$. 

The form of the spectra at redshifts below $z\simeq 2.60$ cannot be obtained within the scheme we are employing, as the
validity of TRG lies within the mildly nonlinear regime. 
At its present development, the scheme cannot account for the process of virialization and formation of
bound structures. As a result, inhomogeneities tend to grow without limit, and the power spectrum diverges. The integration
of eqs.~(\ref{spectev1}), (\ref{spectev2}) below $z=2.60$ results in an increase of the spectrum much faster than the one
predicted by the linear treatment.

\section{Conclusions}
\setcounter{equation}{0}

The numerical analysis of the previous section has led to some concrete conclusions:
\begin{itemize}
\item
The neutrino power spectrum is subdominant to the CDM+BM spectrum
at high redshifts because of free-streaming during the period that
neutrinos are relativistic. The neutrino mass grows during the cosmological evolution of the cosmon field. 
As soon as the neutrinos become nonrelativistic ($z\simeq 8$ in our model), their spectrum grows rapidly and overtakes the 
CDM+BM spectrum. 
\item 
During the period that the neutrinos are nonrelativistic and the evolution linear ($8\gta z \gta 3$),
the neutrino spectrum develops oscillations induced by the BAO in the CDM+BM sector. The oscillations in the neutrino sector
are erased
by the mode coupling generated by the nonlinearities when these become significant ($3\gta z \gta 2.6$).
\item
At high redshifts ($z\gta 3$) the gravitational potentials in the BAO range are dominated by CDM+BM inhomogeneities. 
At lower redshifts, the neutrino inhomogeneities start dominating and they determine the potentials. At the time when
the perturbations in the neutrino sector become highly nonlinear ($z\simeq 2.6$ in our model), the potentials take values in the
 region $[10^{-6}-10^{-5}]$ in the BAO range.
\end{itemize}

The issue that cannot be resolved by our analysis is whether the BAO persist in the CDM+BM sector at redshifts below
the one at which the neutrino sector becomes highly nonlinear. An extrapolation of our results below $z=2.6$ within
a linear treatment would indicate that the potentials induced by the neutrino inhomogeneities grow rapidly extremely large,
so that their influence would erase the BAO from the CDM+BM sector already at redshifts above $z=2$. 
Such a conclusion would make growing neutrino quintessence 
incompatible with observations.
However, the linear increase is strongly misleading, as the neutrino-induced gravitational potential 
would exceed the extreme value
resulting from all neutrinos within the horizon being concentrated at a single point \cite{cmb2}. 
Unfortunately, our present approximation cannot account for a slowing down of the growth as compared to the linear
approximation. 

Nevertheless, an important lesson can be learned from a linear extrapolation. One finds
that, at the time when the oscillatory behavior disappears from
the CDM+BM density spectrum, the gravitational potentials in the BAO range take values in the  region
$[10^{-3}-10^{-2}]$. A significant reduction of the amplitude of oscillations, at the 10--20\% level, corresponds to
potentials in the  region $[10^{-4}-10^{-3}]$. The same critical orders of magnitude of the gravitational potential
persist even if the neutrino spectrum 
is modelled artificially to grow at a much slower rate than the one predicted by an extrapolation of our results: the 
disappearence of BAO at some redshift between $z\sim 2$ and $z\sim 0$ is linked to the presence of gravitational
potentials $\sim 10^{-3}$ or larger. It has been established already that the appearance of 
such potentials would make the predictions of growing neutrino quintessence for the CMB incompatible with its measured 
spectrum,  because of a very strong ISW effect \cite{cmb2}. 

We expect our qualitative conclusions to remain valid for all variations of the scenario of growing neutrino quintessence
(different forms of the potential $U(\phi)$ and the coupling $\beta(\phi)$). The strong influence of the neutrino sector on 
the background evolution requires a large coupling to the cosmon field. In turn, this generates a strong long-range
force between neutrinos. This causes the neutrino perturbations to become nonlinear early in the cosmological evolution,
and the oscillations to disappear from their density power spectrum. The influence on the CDM+BM sector and the 
observable BAO depends on the final size of nonlinear inhomogeneities in the neutrino sector and the 
respective gravitational potentials. Similar considerations would also apply to other models of mass-varying
neutrinos \cite{mvn}, if the neutrinos influence the cosmological evolution considerably.

Our analysis of density spectra in growing neutrino quintessence sets a rough quantitative bound on the
typical gravitational potentials generated by the neutrino inhomogeneities: they should not become much larger than 
$10^{-4}$ in the BAO region. Otherwise, the amplitude of BAO will be reduced  or completely eliminated.
As has been discussed in \cite{collapse,cmb2}, the formation of structures in growing neutrino quintessence is not 
well understood. It is possible that highly nonlinear structures or virialized objects start forming early, and their backreaction 
prevents the neutrino power spectrum from growing very large. It is conceivable that the gravitational potentials  
at large scales remain sufficiently small for the CMB spectrum and the BAO to remain largely unaffected. 
However, a better understanding of structure formation in the scenario of growing neutrino quintessence is required in order to
resolve this issue. As we have mentioned in the introduction, the only approach that seems capable of describing 
the formation of highly nonlinear structures employs hydronamic simulations. In the scenario he have considered,
such simulations become indispensable for the understanding of structure formation already at redshifts $z\sim 3$.

\section*{Acknowledgments}
N.~T. would like to thank M. Pietroni for many useful discussions.
N.~B.  was supported by the EU Marie Curie Network ``UniverseNet'' 
(MRTN--CT--2006--035863).
N.~T. was supported in part by the EU Marie Curie Network ``UniverseNet'' 
(MRTN--CT--2006--035863) and the ITN network
``UNILHC'' (PITN-GA-2009-237920).


\end{document}